# Developing the UML as a Formal Modelling Notation

A. Evans[1], R. France[2], K. Lano[3], B. Rumpe[4]

## Abstract

The Unified Modeling Language (UML) is rapidly emerging as a de-facto standard for modelling OO systems. Given this role, it is imperative that the UML have a well-defined, fully explored semantics. Such semantics is required in order to ensure that UML concepts are precisely stated and defined. In this paper we describe and motivate an approach to formalizing UML in which formal specification techniques are used to gain insight into the semantics of UML notations and diagrams. We present work carried out by the Precise UML (PUML) group on the development of a precise semantic model for UML class diagrams. The semantic model is used as the basis for a set of diagrammatical transformation rules, which enable formal deductions to be made about UML class diagrams. It is also shown how these rules can be used to verify whether one class diagram is a valid refinement (design) of another. Because these rules are presented at the diagrammatical level, it will be argued that UML can be successfully used as a formal modelling tool without the notational complexities that are commonly found in formal specification techniques.

## 1. Introduction

The popularity of object-oriented methods such as OMT [18] and the Fusion Method [4], stems primarily from their use of intuitively-appealing modelling constructs, rich structuring mechanisms, and ready availability of expertise in the form of training courses and books. Despite their strengths, the use of OO methods on nontrivial development projects can be problematic. A significant source of problems is the lack of semantics for the modelling notations used by these methods. A consequence of this is that understanding of models can be more apparent than real. In some cases, developers can waste considerable time resolving disputes over usage and interpretation of notation. While informal analysis, for example, requirements and design reviews, are possible, the lack of precise semantics for OO modelling makes it difficult to develop rigorous, tool-based validation and verification procedures.

The *Unified Modeling Language* (UML) [13] is a set of OO modelling notations that has been standardized by the Object Management Group (OMG). It is difficult to dispute that the UML reflects some of the best modelling experiences and that it incorporates notations that have been proven useful in practice. Yet, the UML does not go far enough in addressing problems that relate to the lack of precision. The architects of the UML have stated that precision of syntax and semantics is a major goal. The UML semantics document (version 1.1) [12] is claimed to provide a "complete semantics" that is expressed in a "precise way" using meta-models and a mixture of natural language and an adaptation of formal techniques that improves "precision while maintaining readability". The meta-models do capture a precise notion of the (abstract) syntax of the UML modelling techniques (this is what meta-models are typically used for), but they do little in the way of answering questions related to the interpretation of non-trivial UML structures. It does not help that the semantic meta-model is expressed in a subset of the notation that one is trying to interpret. The meta-models can serve as precise description of the notation and are therefore useful in implementing editors, and they can be used as a basis to define semantics, but they cannot serve as a precise description of the meaning of UML constructs.

---

[1] Department of Computer Science, The University of York, UK
[2] Department of Computer Science & Engineering, Florida Atlantic University, USA
[3] Department of Computing, Imperial College, London, UK
[4] Department of Computer Science, Munich University of Technology, Germany

The UML architects justify their limited use of formal techniques by claiming that "the state of the practice in formal specifications does not yet address some of the more difficult language issues that UML introduces". Our experiences with formalizing OO concepts indicate that this is not the case. While this may be true to some extent, we believe that much can be gained by using formal techniques to explore the semantics of UML. On the other hand, we do agree that current text-based formal techniques tend to produce models that are difficult to read and interpret, and, as a result, can hinder understanding of UML concepts. This latter problem does not diminish the utility of formal techniques, rather, it obligates one to translate formal expressions of semantics to a form that is digestible by users of the UML notation.

In a previous paper [8], we discuss how experiences gained by formalizing OO concepts can significantly impact the development of a precise semantics for UML structures. We motivated an approach to formalizing UML concepts in which formal specification techniques are used primarily to gain insights tothe semantics of UML notations. In this paper we present the roadmap we are using to formalize the UML, and describe the results of its application to the formalization of UML static models.

The primary objective of our work is to produce rigorous development techniques based on the UML. A first step is to make UML models amenable to rigorous analyses by providing a precise semantics for the models. This paves the way for the development of formal techniques supporting the rigorous development of systems through the systematic enhancement and transformation of OO models. In this paper we show how the formalized static model can be rigorously manipulated to prove properties about them and their relationships to other static models.

In section 2, we give an overview of work on the formalization of OO modelling concepts and notations, and outline the PUML formalization approach. In section 3 we present the results of our work on formalizing UML static models, and in section4 we show how the resulting static model diagrams can be formally manipulated.

We conclude in section 5 with a summary and a list of some of the open issues that have to be tackled if our approach is to bear meaningful results.

## 2. Formalizing OO Concepts: Overview and Roadmap

**Classification of Approaches**

In [8] we identified three general approaches to formalizing OO modelling concepts: *supplemental*, *OO-extended formal notation*, and *methods integration* approaches. In the supplemental approach more formal statements replace parts of the informal models that are expressed in natural language. Syntropy [5] uses this approach. In the OO-extended formal language approach, an existing formal notation (e.g. Z [20]) is extended with OO features (e.g. Z++ [17] and Object-Z [ 6]). In the methods integration approach informal OO modelling techniques are made more precise and amenable to rigorous analysis by integrating them with a suitable formal specification notation (e.g., see [9,2,14]).

Most method integration works involving OO methods focus on the generation of formal specifications from less formal OO models. This is in contrast to the PUML objectives, where the OO models are the precise (even formal) models. The degree of formality of a model is not necessarily related to its form of representation. In particular, graphical notations can be regarded as formal if a precise semantics is provided for their constructs.

A formal semantics for a modelling notation can be obtained by defining a mapping from syntactic structures in the (informal) modelling domain to artifacts in the formally defined semantic domain. This mapping, often called a meaning function, is used to build interpretations of the informal models.

Rather than generate formal specifications from informal OO models and require that developers manipulate these formal representations, a more workable approach is to provide formal semantics for graphical modelling notations and develop rigorous analysis tools that allow developers to directly manipulate the OO models they have created. Defining meaning functions provides opportunities for exploring and gaining insight into appropriate formal semantics for graphical modelling constructs.
The method developers (and not the application developers) should use these mappings to justify the correctness of analysis tools and procedures provided in a CASE tool environment.

**Roadmap to Formalization**

Our experiences with formalizing OO modelling notations indicate that a precise and useful semantics must be complete (i.e., meanings must be associated with each well-formed syntactic structure), preserve the intended level of abstraction
(i.e., the elements in the semantic domain must be at the same level of abstraction as their corresponding modelling concepts), and understandable by method developers. Furthermore, the formalization of a heterogeneous set of modelling techniques requires that the notations be integrated at the semantic level. Such integration is required if dependencies across the modelling techniques are to be defined.

The following are the steps of the formalization approach that we use in our work on formalizing the UML:

1. In this step, a formal language for describing syntax and semantics is chosen. For the UML formalization we chose Z because it is a mature, expressive and abstract language, that is well supported by tools. Our experiences with using Z to formalize OO concepts indicates that it is expressive enough to characterize OO concepts in a direct manner (i.e., without introducing unwanted detail).

2. In this step, the abstract syntax of the graphical OO notation is defined. Here, we will refer to this notation as (language) L. Language L, like conventional textual languages, needs to have a precise syntax definition. Whereas grammars are well suited for text, the UML meta-model [11] works well as a description of the structure of UML diagrams. However, a Z characterization of the abstract syntax is better able to capture constraints on the syntactic structures that can be formed using the graphical constructs.

3. This step is concerned with characterizing the notion of a system in terms of its constituent parts, interactions, and static and behavioral properties. The characterization defines the elements of the semantic domain, which we denote by S. The elements of the semantic domain correspond to modelling concepts that are independent of particular modelling techniques. In the OO modelling realm this is possible because objects have certain properties that are independent from the modelling techniques, and are thus intrinsic to "being an object". In [16] and [19] a *system model* is defined, and used as the semantic domains for OO notations in papers such as [3] and [19]. In this paper, the semantic domain is characterized using the language Z.

4. This step is concerned with defining the meaning function for the OO notation. A mapping between the syntactic domain L and the semantic domain S is defined. The system model domain formally defines the set of all possible systems. The semantics of a model created using a given description technique is obtained by applying the meaning function to its syntactic elements. The semantics of a model is given by a subset of the system model domain. This subset of the system model consists of all the systems that possess the properties specified in the model.

5. In the final step, analysis techniques are developed for the formalized OO notation. These techniques enable us to constructively enhance, refine and compose models expressed in the language L, and also allow us to introduce verification techniques at the diagrammatic level.

An important aspect of our formalization approach is the separation of concerns reflected in the language-independent formulation of the semantic domain S. This leads to a better understanding of the developed systems, allows one to understand what a system is independently of the used notation, and allows one to add and integrate new OO diagramming forms.

Though we speak of one language L, this language can be heterogeneously composed of several different notations. However, it is important to note that integration of these notations is more easily accomplished if the semantic domain S is the same for all these sub-languages.

In the following sections, we illustrate the application of this formalization approach using a small subset of UML class diagram notation.

## 3. A Formalization Example

In this section we formally define the abstract syntax of a subset of the UML static model notation, characterize an appropriate semantic domain for its components, and define a meaning function for the formally defined syntax.

**Abstract Syntax**

In the UML semantics document (version 1.1), the core package - *relationships* - gives an abstract syntax for the static components of the UML. This is described at the meta-level using a class diagram with additional well-formedness rules given in OCL. For reasons given in the previous section, we use the Z notation to define the abstract syntax. Unlike the OCL, Z provides good facilities for proof. In our work we treat the UML semantics document as a requirements statement from which a fully formal model can be obtained.

As an example, the following schemas define some of the UML static model constructs. Specifically, they define a set of classifiers, associations and a generalization hierarchy, and attach a set of attributes to each classifier:

$$[Classifier, Name]$$

An association end connects an association to a classifier, and has a unique name and multiplicity:

```
─ AssociationEnd ──────────────
  name : Name
  classifier : Classifier
  multi : ℙ ℕ
───────────────────────────────
```

Each association is connected to a number of association ends:

```
─ Association ─────────────────
  name : Name
  connects : 𝔽 AssociationEnd
───────────────────────────────
```

The components of the abstract syntax are as follows:

```
─ Static1 ─────────────────────────────────
  abstract, classifiers : 𝔽 Classifier
  associations : 𝔽 Association
  attributes : Classifier ⇸ 𝔽 Name
  supertype_of : Classifier ↔ Classifier
  ─────────────────────────────────────────
  abstract ⊆ classifiers
───────────────────────────────────────────
```

Well-formedness of the abstract syntax is ensured by further constraints:

```
┌─ Static ─────────────────────────────────────────────────────────
│ ┌─ Static1 ──────
│ │
│ │ supertype_of ∈ (classifiers ↔ classifiers)
│ │ supertype_of⁺ ∩ id(classifiers) = ∅
│ │ ∀ c₁, c₂ : classifiers •
│ │     c₁ supertype_of c₂ ⇒ attributes(c₂) ⊆ attributes(c₁)
│ │ ∀ a₁, a₂ : associations •
│ │     a₁ ≠ a₂ ⇒ a₁.name ≠ a₂.name
│ │ ∀ a : associations •
│ │     {e : a.connects • e.classifier} ⊆ classifiers
```

The above schema describes the constraints governing how elements of the abstract syntax can be combined (more constraints are possible). These constraints state that:

- the collection of classifiers in the supertype hierarchy form a directed acyclic graph;
- associations are unique and link a classifier to another classifier (or to itself).

**Semantic Domain**

Semantically, a classifier is represented as a set of objects. Each instance has a unique value, which distinguishes it from all other object and non-object values:

$$[Value]$$

```
┌─ Values ───────────────────────────
│ oValues, nValues : ℙ Value
│ ─────────────────────
│ disjoint ⟨oValues, nValues⟩
```

An object is owned by a classifier, has a unique identity, and maps a set of attributes to values:

```
┌─ Object ────────────────────
│ classifier : Classifier
│ self : Value
│ attributes : Name ⇸ Value
```

At any point in time, a system can be described as a set of objects, where each object is referenced by it's identity self:

```
┌─ SM1 ─────────────────────────────────────────────
│ Values
│ objects : Value ⇸ Object
├───────────────────────────────────────────────────
│ dom objects = oValues
│ ∀ o : Value • (objects(o)).self = o
└───────────────────────────────────────────────────
```

From that snapshot, we can derive sets of links (instances of associations):

```
┌─ SM ──────────────────────────────────────────────
│ SM1
│ links : Name → (Value ↔ Value)
├───────────────────────────────────────────────────
│ ∀ at : Name; o₁, o₂ : Value •
│     (o₁, o₂) ∈ links(at) ⇔ o₂ = (objects(o₁)).attributes at
└───────────────────────────────────────────────────
```

**Semantic Mapping**

The semantic mapping determines how the syntactic elements of the UML static model, for example, abstract, classifier, and association, are to be interpreted in the semantic domain. The semantic mapping that takes the concepts given in the syntactic domain AbstractSyntax to elements in the semantic domain SM is characterized by a Z schema that takes the characterizations of the syntactic and semantic domains as parameters.

```
┌─ Semantics ───────────────────────────────────────
│ Static
│ SM
├───────────────────────────────────────────────────
│ {o : ran objects • o.classifier} ⊆ classifiers \ abstract
│
│ ∀ o : dom objects •
│     attributes((objects(o)).classifier) ⊆ dom(objects(o)).attributes
│
│ ∀ a : associations; o : dom objects •
│     ∀ e : a.connects •
│         e.classifier = (objects(o)).classifier ⇒
│             e.name ∈ dom(objects(o)).attributes ∧
│             #((links(e.name))⦇{o}⦈) ∈ e.multi
│
│ ∀ s₁, s₂ : Classifier •
│     s₁ supertype_of s₂ ⇔
│         {o : Value | (objects(o)).classifier = s₂} ⊆
│             {o : Value | (objects(o)).classifier = s₁}
└───────────────────────────────────────────────────
```

The axioms state that each object is assigned to a non-abstract classifier. Furthermore, the objects have at least the set of attributes explicitly mentioned in the classifier definitions. We also interpret association ends as attributes and restrict the multiplicities. Finally, the supertype relationship requires that set of objects assigned to a subtype is a subset of the objects assigned to its supertype.

An explicit form of the meaning function can be expressed as follows:

$$M : Static \to \mathbb{P}\, SM$$
$$\forall st : Static \bullet$$
$$M(st) = \{Semantics \mid st = \theta Static \bullet \theta SM\}$$

## 4. Analyzing UML diagrams

As discussed above, a central part of the PUML group's work is to develop a formal version of UML that can be used to build precise and analyzable models. However, how can a UML model be analyzed? In the case of a textual notation such as Z, analysis is carried out by constructing *proofs* to determine the truth or falsity of some property being asserted about a specification. Each proof involves applying a sequence of inference rules and axioms to the specification to derive the required conclusion. At each step, a new formula is derived either from the original specification or as a result of applying an inference rule to previous formulas.

To analyze UML models, a very similar approach can be adopted [7]. However, because UML is a diagrammatical modelling language, a set of deductive rules for UML will consist of a set of *diagrammatical transformation rules*. Thus, proving a property about a UML model will involve applying a sequence of transformation rules to the model diagrams until the desired conclusion is reached. As an example, consider a class diagram, which describes the relationship between a university and its students. If a student can be specialized as being either part-time or full-time, can it be deduced (by suitable transformations) that the university has the same relationship with a full-time student as it has with all students?

The following diagrams can express this (obviously correct) theorem. Here the diagram on the right expresses the theorem to be proved:

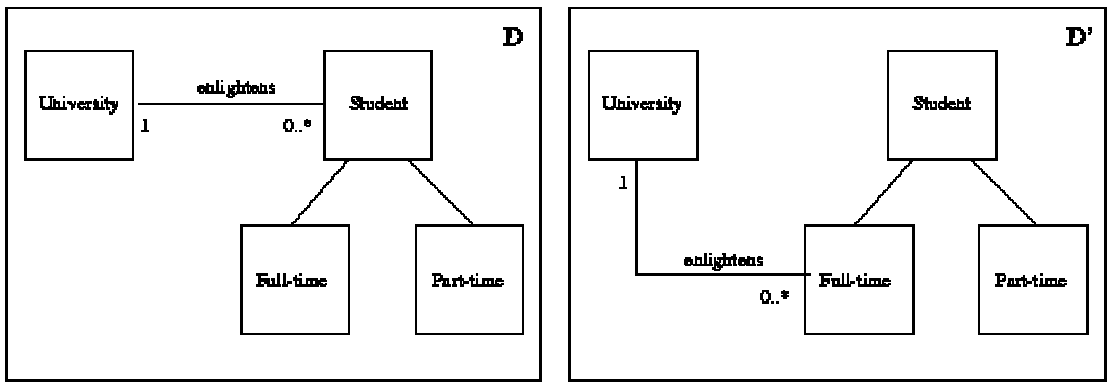

Using a suitable sequence of transformation rules, we should be able o transform the original diagram into the second diagram, thereby roving that the theorem is valid. In this case, three steps are required to carry out the proof, each of which is the result of applying a specific transformation rule. The first step is to introduce a new association (e.g. new) between University and Part-time as a weakened version of enlightens. The second step is to erase the original association enlightens. The proof is completed by renaming new to enlightens.

Analysis rules offer an intuitive method of reasoning with UML models. In addition, they have a number of other important applications:

**Refinement proofs**: UML analysis rules can be used to prove that one model is a refinement of another. Given two models (or diagrams) M' and M, then we say that M' is a refinement of M, if M can be deduced from M'. Thus, any property that holds for the concrete model M' must also hold for the abstract model M (but not necessarily the reverse). As an example, consider the diagrams shown above. Because we can show that the second diagram D' is a deduction from the first diagram D, then it must also be true that the first diagram is a refinement of the second diagram. This conclusion seems to match our intuitive notion of refinement as a process of strengthening assumptions made about a model. In this case, we have chosen to strengthen the relationship between the University and its students by choosing to require that *all* students must be enlightened, rather than just full-time students!

Please note, deduction is a technique to derive properties that are already given more or less implicitly within a model. Instead, refinement focuses on adding new properties of to a model, thus enhancing it. These two techniques are the exact opposite and therefore, deduction rules applied in the opposite direction lead to refinement rules.

**Design pattern verification**: a further use for the analysis rules might be in the verification of design patterns [10]. A design pattern is just an example of a transformation on a static or dynamic model, which is refinement preserving. However, at present, most design patterns are not proven correct, and are therefore open to misuse and incorrect definition. Analysis rules, in combination with a sound semantic base, are a means by which this problem can be overcome.

Whenever a transformation rule is applied to a diagram it must be shown that the resulting diagram is a valid deduction of the original diagram. The condition under which this is true is known as the *satisfaction* condition. This states that if every meaning satisfying one model also satisfies another model, then whatever property holds for the first model must also hold for the second. Thus, the second diagram follows from (or is a logical deduction of) the first diagram. Of course, for this result to be valid, both models must be well formed.

This condition can be expressed in Z as follows: Let us assume, there is a transformation rule T given. This is formally represented as a modification on the syntax, in this case a static model:

$$T : Static \twoheadrightarrow Static$$

Such a transformation, can, for example, be the addition of a new classifier, the specialization of a multiplicity, or the join of several static models. This syntactic transformation needs a semantic counterpart, which relates elements of the semantic domain. This is known as the satisfaction relation, and it has the general form:

$$\_ \models \_ : \mathbb{P}(SM) \leftrightarrow \mathbb{P}(SM)$$

The satisfaction relation forms the basis for unambiguously defining the conditions under which a diagram can be considered to *satisfy* the properties of another diagram. Defining suitable satisfaction conditions for this relation will be an essential step towards our aim of developing rigorous analysis methods for UML. For example, one possible satisfaction relation might permit both introduction and deletion of classes as a deductive step, whilst another might only permit class introduction[5].

Finally, the formal proof of correctness of a transformation can now be described within Z (and therefore can be proven within Z). A transformation T is correct, if

$$\forall st : AbstractSyntax \bullet \mathcal{M}(st) \models \mathcal{M}(T(st))$$

This strongly corresponds to the commuting diagram, first stated in [19] and also in [15].

### 5. Summary and Open Issues

In this paper we outlined and illustrated an approach to formalizing the UML. The objective of our efforts is to make the UML itself a precise modelling notation so that it can be used as the basis for a rigorous software development method. However, it must first be determined how such a formalization can best be carried out, and what practical purpose it can serve. This paper aims to contribute to this ongoing discussion.

The benefits of formalization can be summarized as follows:

- Lead to a deeper understanding of OO concepts, which in turn can lead to more mature use of technologies.

- The UML models become amenable to rigorous analysis. For example, rigorous consistency checks within and across models can be supported.

- Rigorous refinement techniques can be developed.

An interesting avenue to explore is the impact a formalized UML can have on OO design patterns and on the development of rigorous domain-specific software development notations. Domain-specific UML patterns can be used to bring UML notations closer to a user's real-world constructs. Such patterns can ease the task of creating, reading, and analyzing models of software requirements and designs.

An integrated approach to formalization of UML models is needed in order to provide a practical means of analyzing these models. Current work on compositional semantics [1] has used techniques for theory composition to combine semantic interpretations of different parts of an OO model set.

---

[5] At present we are investigating a number of different satisfaction relations for UML diagrams in order to determine which best fits emerging practice.

Some of the other issues that have to be addressed in our work follows:

- How does one gauge the appropriateness of an interpretation of UML constructs? In practice an `accepted' interpretation is obtained by consensus within a group of experts. Formal interpretations can facilitate such a process by providing clear, precise statements of meaning.

- Should a single formal notation be used to express the semantics for all the models? The advantage of a single notation is that it provides a base for checking consistency across models, and for refinement of the models. This is necessary if analysis and refinement is done at the level of the formal notation. On the other hand, if the role of the formal notation is to explore the semantic possibilities for the notations, and analysis and refinement are carried out at the UML level, then there seems to be no need to use a single formal notation.

It is anticipated that as our work progresses additional issues that will have to be tackled will surface.